\documentclass{aa}
\usepackage{txfonts}
\usepackage{graphicx}
\begin{document}
\title{The secrets of T~Pyx: I. UV observations \thanks{Based on
observations made with the International Ultraviolet Explorer and
de-archived from the ESA VILSPA Database. }} 
\author{Roberto Gilmozzi \inst{1} \and Pierluigi Selvelli \inst{2} }
\offprints{R. Gilmozzi}
\institute{European Southern Observatory, Karl-Schwarzschild-Str. 2,
D-85748 
Garching, Germany \\
\email{rgilmozz@eso.org} 
\and INAF-Osservatorio Astronomico di Trieste, Via Tiepolo 11,  
I-34143  Trieste,  Italy\\
\email{selvelli@oats.inaf.it}} 

\date{Received.....; accepted }

\abstract
{} {We study the UV spectral behavior of the recurrent nova T~Pyx
during 16 years of \sl IUE \rm observations.}  
{We examined both the
\sl IUE \rm  line-by-line images and the extracted spectra in order to
understand the reality and the origin of the observed spectral
variations. We compare different extraction methods and their
influence on the spectrum of an extended object.} {The UV continuum of
T~Pyx has remained nearly constant in slope and intensity over this
time interval, without any indication of long-term trends. The
reddening determined from the UV data is $E_{B-V}=0.25 \pm 0.02$. The
best single-curve fit to the dereddened UV continuum is a power-law
distribution $\propto \lambda^{-2.33}$. The tail of this curve agrees
well with the B, V, and J magnitudes of T~Pyx, indicating that the
contribution of the secondary star is negligible.  One peculiar aspect
of T~Pyx is that most emission lines (the
strongest ones being those of CIV $1550$ and HeII $1640$) show
substantial changes both in intensity and detectability,  in contrast to
the near constancy of the continuum. Several
individual spectra display emission features that are  difficult
to identify, suggesting a composite spectroscopic
system. We tentatively ascribe the origin of these transient emission
features either to loops and jets from the irradiated secondary or to
moving knots of the surrounding nebula that are (temporarily)
projected in front of the system. The inspection of all IUE
line-by-line images has led to the detection of emission spikes
\it outside \rm the central strip of the spectrum, which in some cases
seem associated to known emission features  in the (main)
spectrum.  A comparison with other ex-novae reveals a surprising
similarity to the spectrum of the very-slow nova HR Del, whose white
dwarf primary has a mass that is allegedly about one half that of
T~Pyx.} {}

\keywords{Stars: novae -- Ultraviolet: stars -- Stars: recurrent novae}

\maketitle

%
%

\section{Introduction}

The five recorded outbursts of the recurrent nova (RN) T~Pyx occurred
in 1890, 1902, 1920, 1944, and 1966, with a mean recurrence time of
19$\pm$5.3 yrs (Webbink et al., 1987, hereinafter WLTO).  All
outbursts were remarkably similar in photometric behavior and
characterized by a speed class that was substantially slower than in
other RNe, with t$_2$ = 60$^{\rm d}$ instead of a few days.
We note that  according to the classifications of
Payne-Gaposchkin (1957) and Warner (1995) T~Pyx should be considered
``moderately fast" and not ``extremely slow" as reported in previous
works.

It is unfortunate that the orbital period of the binary system has not
been clearly established. The published results range from values as low
as
1$^{\rm h}$ 40$^{\rm m}$ (Szkody and Feinswog 1988, from a single sine
solution of J photometry) to values as high as 3$^{\rm h}$ 29$^{\rm
m}$ (Vogt et al., 1989, from a preliminary study of spectroscopic data
with limited spectral resolution).  A more recent and comprehensive
photometric study by Schaefer et al. (1992) based on data from 114
nights of observations, together with measurements from the literature,
has led to the disappointing conclusion that T Pyx has an incoherent
photometric period and that the sinusoidal modulation is not closely
tied to the orbital period. By analogy with other stars that have
variable photometric periods (possibly related to the superhump
phenomenon),  Shaefer et al. (1992) suggested an orbital period of near
1$^{\rm h}$ 45$^{\rm m}$ from the ``best" period for the modulation
(1$^{\rm h}$
49$^{\rm m}$).  The relation  between the orbital periods of
cataclysmic variables and the spectral types of the secondary stars
(Smith and Dhillon 1998) indicates  that  the secondary star in T Pyx
is  probably  an  M4-5 dwarf.

Already in 1979 ground-based imaging by Duerbeck and Seitter (1979)
detected a shell of ejected matter, a few arcsec across, that surrounded
the
system.  Subsequent observations by Seitter (1986) and Duerbeck (1987)
revealed the presence of structures in the shell, while Shara et al.
(1989) discovered the additional presence of a fainter extended shell
with a radius of about 10 arcsec.  Thanks to the high spatial resolution
of the WFPC2 on board the HST, the ``shell" has been resolved into
more than two thousand individual knots (Shara et al. 1997). 

These observations have provided  direct evidence that the ejecta
of novae, previously believed to be reasonably homogeneous, are instead
highly structured, confirming the GHRS observations by Shore et al.
(1993) that established the presence of knots in Nova Cyg 1992 and the
indirect conjectures on the existence of knots previously made by
Krautter et al. (1984), Williams et al. (1991), and Saizar and Ferland
(1994). 

Shabbaz et al. (1997) interpreted the presence of two spectral features
on either side of H$\alpha$ as the signature of a bi-polar,
high-velocity, highly collimated jet, but subsequent studies by Margon
and Deutsch (1998) and by O'Brien and Cohen (1998) have shown that these
features are instead emission in the [NII] 6548 and 6584 lines
from a complex velocity field in the surrounding shell.

\subsection{The missing outburst}

Observations of an RN for which an outburst is expected to occur
provide a unique opportunity for a direct observational test of the
theoretical expectations of the thermonuclear runaway (TNR) theory.
From the observed $\dot{M}$ and the inter-outburst interval, one can
indeed estimate the total accreted mass $M_{\rm accr}$ and compare it
to both the theoretical mass for ignition $M_{\rm ign}$ and the
observed mass of the ejected shell $M_{\rm ej}$. Ultraviolet
observations  are
best-suited for these studies since most of the accretion disk
luminosity is emitted in the far-UV and many emission lines of
important astrophysical species are present in this wavelength range.

With this rationale in mind and in view of the allegedly imminent
outburst of T~Pyx,  as expected on the basis of its apparently
regular behavior, with a  mean interval of
about 19 years between  
the observed outbursts (WLTO), and of the about twenty years
elapsed from the last outburst of December 1966, we started  an
observing program in 1986 with {\sl
IUE} with the purpose of monitoring the quiescent phase and to follow
the spectroscopic changes in the phases prior to the outburst and
during the early outburst. Actually, the star has
successfully managed  to  postpone the long-awaited outburst, and
at the present time (2006)
has surpassed by sixteen years the longest inter-outburst interval so
far recorded (24 yrs).

Shortly before this paper neared completion, Schaefer (2005) 
published the results of a study on the interoutburst interval in
recurrent novae. From an analysis of the data on the magnitudes
obtained from archival plates, from the literature, and from his own
collection of CCD magnitudes, he found that the product of the
inter-eruption interval times the average bolometrically corrected
flux is a constant for both T Pyx and U Sco. This finding, together
with the apparent decline in the observed B magnitude in the last
inter-eruption intervals (Fig. 1 in Shaefer's paper), has provided a
physical basis for predicting that the next outburst of T Pyx will
occur around 2052.

While waiting to (hopefully) personally verify these predictions, we
present and discuss here (Paper I) the main spectroscopic results from
16 years of {\sl IUE} observations.  In a follow-up paper (Selvelli
and Gilmozzi 2006, Paper II), we will combine the results of these
observations with other basic information on the system (including the
implications of the findings by Schaefer (2005) on the accretion rate
of T Pyx during the past inter-eruption intervals), in order to set
stringent constraints on the physical parameters of this elusive object
and to refine our understanding of the recurrent nova phenomenon.

\section { The IUE observations and data reduction}

The only {\sl IUE} observations of T~Pyx previous to ours were those
of May 11, 1980 by Seitter and Duerbeck and reported by Bruch et al.
(1981).  In this preliminary study they found $E_{B-V}\sim 0.35$ and
gave an estimate for the distance of about 2500 pc.

The {\sl IUE} spectra cover the time interval from May 11, 1980 to May
7, 1996 (with a gap between 1980 and 1987).  In total, 37 short
wavelength (SWP camera) and 14 long wavelength (LWP camera) spectra
are available, although some SWP spectra present a low S/N  because
of high background noise and/or a bad photometric response as a
consequence of unusual centering in the {\sl IUE} aperture.  Typical
exposure times for the SWP spectra were around three hours with a few
exposures twice as long (Table 1).

An important aspect of data reduction is that the {\sl IUE}
data extraction and calibration methods have undergone several
revisions during and after the {\sl IUE} lifetime, which has led to a
general improvement in the quality of the line spectrum.  While this
work was in progress (from 1987 to the last spectra secured in May
1996), both the IUESIPS and then NEWSIPS extraction procedures were
adopted.  See Gilmozzi et al. (1998) for some considerations of the
difference in the quality of data extraction using IUESIPS or NEWSIPS.

The spectra used in this paper  were instead retrieved from the
INES ({\sl IUE} Newly Extracted Spectra) final archive.  The more
relevant modifications in the INES system,  in comparison with 
NEWSIPS,
are: 1) the adoption of a new noise model, 2) a more accurate
representation of the spatial profile of the spectrum, 3) a more
reliable determination of the background, 4) a more adequate treatment
of ``bad" pixels, and 5) the improvement in the handling and
propagation of the quality flags to the final extracted spectra.  In
general, the extracted fluxes of INES and NEWSIPS are in excellent
agreement, but especially for spectra with weak continua and narrow
emission lines the INES extraction can register significantly more
flux (up to ten sigma!)  (Schartel and Skillen 1998).  For a detailed
description of the {\sl IUE}-INES system see Rodriguez-Pascual et al.
(1999) and Gonzalez-Riestra et al.  (2001).

\begin{table}
\caption{{\sl IUE} observation  log}
\centering
{\tiny
\begin{tabular}{llcrl}
Camera &  Image  &  Date & ExpTime [s] & Comments \\
LWR &  07724  & 1980-05-11 &   7199.8   & \\
SWP &  08973  & 1980-05-11 &  12899.8  & \\
LWP &  09204  & 1986-09-27 &   7199.8   & \\
SWP &  29318  & 1986-09-27 &  16199.5  & \\
SWP &  32218  & 1987-11-02 &  16799.6  & \\
LWP &  11996  & 1987-11-02 &   6599.7   & \\
LWP &  12644  & 1988-02-11 &   7199.8   & \\
SWP &  32899  & 1988-02-11 &  12779.8  & \\
LWP &  12791  & 1988-03-03 &  12779.8  & \\
SWP &  33034  & 1988-03-04 &  23579.7  & \\
LWP &  14383  & 1988-11-05 &   7799.4   & \\
SWP &  34696  & 1988-11-05 &  17159.6  & \\
LWP &  16757  & 1989-11-07 &   7799.4   & \\
SWP &  37536  & 1989-11-07 &  16799.6  & \\
SWP &  43442  & 1991-12-22 &  15599.4  & \\
LWP &  22052  & 1991-12-22 &   7199.8   & \\
SWP &  44182  & 1992-03-16 &  14999.8  & \\
LWP &  22608  & 1992-03-16 &   9599.6   & \\
SWP &  44948  & 1992-06-17 &  16799.6 &  \\
LWP &  23317  & 1992-06-18 &   6899.5   & \\
LWP &  24612  & 1992-12-28 &   8399.5   & \\
SWP &  46605  & 1992-12-28 &  16319.5  & \\
SWP &  47057  & 1993-02-27 &  16499.7  & \\
LWP &  25020  & 1993-02-27 &   6599.7   & \\
SWP &  47323  & 1993-03-20 &  20099.7  &   poor center\\  
SWP &  47328  & 1993-03-21 &  17999.7  &   poor center\\
SWP &  47332  & 1993-03-22 &  23099.6  & \\
SWP &  49365  & 1993-11-29 &   9599.6   &   noisy\\  
SWP &  49366  & 1993-11-29 &  10679.7  &   noisy\\
SWP &  50099  & 1994-02-24 &  11399.8  & \\
SWP &  50100  & 1994-02-24 &  10799.7  &   noisy \\
SWP &  50596  & 1994-04-20 &  24479.6  & \\
SWP &  52886  & 1994-11-23 &   8999.6   & \\
SWP &  52887  & 1994-11-23 &   9479.6   &   noisy\\ 
SWP &  53809  & 1995-02-02 &  11399.8  & \\
SWP &  53810  & 1995-02-02 &  10679.7  &   noisy\\ 
SWP &  54590  & 1995-05-03 &  11099.6  & \\
SWP &  54591  & 1995-05-04 &  11699.6  &   noisy\\  
SWP &  56240  & 1995-11-26 &  11999.5  & \\
SWP &  57030  & 1996-05-01 &  11999.5  & \\
SWP &  57031  & 1996-05-02 &  12599.5  & \\
SWP &  57032  & 1996-05-02 &  11999.5  & \\
SWP &  57033  & 1996-05-03 &  11999.5  & \\
SWP &  57034  & 1996-05-03 &  10799.7  & \\
SWP &  57035  & 1996-05-03 &   6599.7   & \\
SWP &  57039  & 1996-05-04 &   7799.4   & \\
SWP &  57042  & 1996-05-04 &   9899.4   & \\
SWP &  57047  & 1996-05-05 &  10799.7  & \\
LWP &  32286  & 1996-05-05 &   5699.8   & \\
SWP &  57055  & 1996-05-06 &  10799.7  & \\
LWP &  32287  & 1996-05-06 &   6599.7   & \\
\end{tabular}
}
\end{table}

\begin{figure}
\centering
\resizebox{\hsize}{!}{\includegraphics[angle=-90]{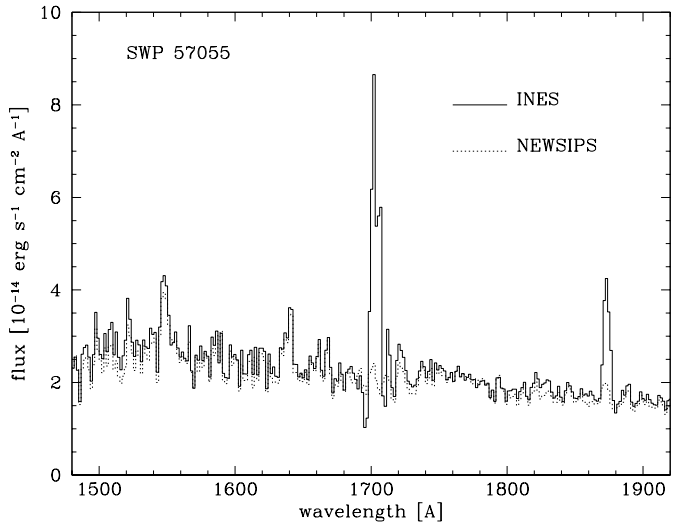}}
\caption{Comparison of NEWSIPS and INES extraction methods. Note the two
strong lines near 1703 and 1873 \AA~ present only in the INES
extraction.}.
\end{figure}

For T~Pyx, a comparison between the NEWSIPS and INES extraction
methods of the {\it same} spectrum has shown in several cases
remarkable differences in the strength of some emission
features. Figure 1 (SWP57055) is an outstanding example of this problem
that may surely help to emphasize the difficulties we have had in
assessing the reality of some spectral features.  The features at 1703
and 1873 \AA~ in the INES extracted spectrum are due to knots of
emission located outside the strip that defines the spectrum (they are
flagged as ``cosmic rays" but  that they are quite wide and
both lie at exactly the same distance from the spectrum might indicate
that they may be real spectral features originating from an extended
region, see also Fig. 6).  While NEWSIPS  automatically removes these
lines (with the risk of removing also real emission features), INES
is more conservative and keeps the features but puts a proper quality
flag in the corresponding spectrum column.  In view of
these problems that are especially critical when one deals with the
spectrum of an object that is surrounded by an extended shell, we have
carefully
checked all line-by-line (SILO) images for the presence of similar
pseudo-emission features  in
order to guarantee the reality of the emission features that are
present in the various INES extracted spectra (see also Sect. 4.2).

Another non minor consequence of the migration from IUESIPS to NEWSIPS
and finally to INES is in the effects that the changes in the
extraction methods and in the absolute calibration tables have had on
the shape of the continuum and consequently on the depth of the
interstellar absorption bump.  In turn, this has resulted in non
negligible changes in our estimates of the reddening and of the
UV-integrated flux and consequently on the distance, the UV
luminosity, and ultimately the mass accretion rate (see Paper II).

\section {The continuum energy distribution}

The observed UV continuum distribution of T~Pyx has remained nearly
constant in slope and intensity over the 16 years of {\sl IUE}
observations showing only minor variations and no indication of any
long term trends.  In the SWP region the standard deviation is only
3.7\%, even including the badly exposed spectra.  This has justified
the creation of an average spectrum (Fig. 2, top) by co-adding and
merging all SWP and LW spectra that are not underexposed or badly
centered.

\begin{figure}
\centering
\resizebox{\hsize}{!}{\includegraphics{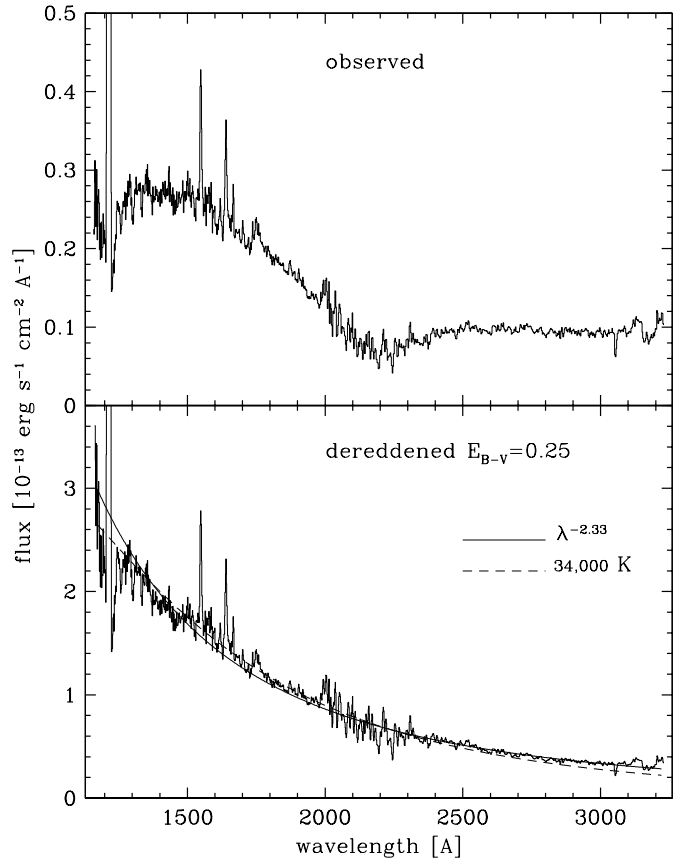}}
\caption{Average {\sl IUE} spectrum of T~Pyx obtained by co-adding and
merging 35 SW and 14 LW  IUE spectra. {\it Top:} observed, {\it
Bottom:} dereddened with $E_{B-V}=0.25$ (this paper). }.
\end{figure}

The improved S/N in the average spectrum has allowed both the
detection of weak line features and an accurate determination of the
reddening.  We obtained $E_{B-V}=0.25 \pm 0.02$ by the standard
method of removing the broad interstellar extinction feature centered
at 2175 \AA~ using the average interstellar absorption curve of Savage
and Mathis (1978).  Therefore in the present study and in Paper II we
 adopt A$_{\rm v}$=3.15$\times$0.25$\sim$0.78.  The same method
applied earlier to the smaller sample of spectra processed with
IUESIPS gave $E_{B-V}=0.31$, while for NEWSIPS spectra it gave
$E_{B-V}=0.24$.

One should note that there is some confusion in the literature about
the $E_{B-V}$ value for T~Pyx: WLTO used the preliminary value
$E_{B-V}$=0.35 by Bruch et al. (1981, but they reported it as 0.36)
and therefore obtained dereddened colors $(B-V)_{\rm o} = -$0.26 and
$(U-B)_{\rm o} = -$1.25, which look too negative. These same colors were
 subsequently adopted by Patterson et al. (1998), who on this basis
concluded that ``T~Pyx is the bluest nova remnant in the sky".  Bruch
and Engel (1994)  reported $E_{B-V}$=0.20 and (B-V)=0.14 from
which they derived $(B-V)_{\rm o}=-$0.06.  (This $E_{B-V}$=0.20 value
is probably  a typo, because they give Bruch et al. 1981 as a
reference, but, unfortunately, this same value was listed also by
Szkody, 1994).  Finally, Weight et al. (1994)  gave $E_{B-V}$=0.08 
in a near-IR study of old classical novae.

\begin{figure}
\centering
\resizebox{\hsize}{!}{\includegraphics[angle=-90]{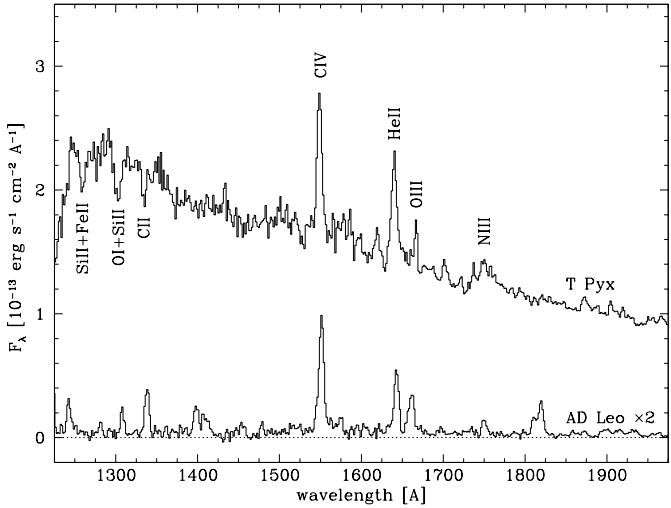}}
\caption{Details of the dereddened average spectrum of T Pyx in the SW
range.  The spectrum of the late-type active star AD-Leo (bottom) is
included for comparison (see Section 6.2).}
\end{figure}

The UV continuum is clearly defined longwards of 1400 \AA~ where only
a few and well-known emission lines are present in the SWP spectra,
while the LW region is poorer in line features.  For a correct
positioning of the continuum in the region below 1400 \AA, where it
was not clear whether the spectral features were actually in
absorption or in emission, we have been helped by a preliminary
identification of these line features (most of them come out as
absorption lines, see Sect. 5).  The presence of a wide interstellar or
circumstellar Ly$\alpha$ absorption, which extends longward up to 1250
\AA, together with that of the strong Ly$\alpha$ emission of
geocoronal origin has also made it rather difficult to estimate the
continuum below 1260 \AA.  Therefore, we  first determined the
continuum in the range 1260-3200 \AA~ using only line-free regions,
fitted it with a spline, and extrapolated it to shorter wavelengths.
A less interactive fitting method (the {\sl icfit} task in IRAF) 
yielded a very similar result.

The best single-curve fit to the UV continuum is a power-law
distribution $F_{\lambda}=4.28\ 10^{-6}\ \lambda^{-2.331}$ (erg
cm$^{-2}$s$^{-1}$ \AA$^{-1}$), with a small uncertainty of $\pm$0.04
in the index. This spectral index is exactly the Lynden-Bell law for a
standard disk ($\alpha=-2.33$). A black body with $T=34160 K$ is also
a good fit, although it would be difficult to ascribe a single
temperature to the most  plausible source of the continuum, the
accretion
disk.  The lower half of Fig. 2 shows a plot of the dereddened average
spectrum
together with the continuum fits (see also Fig. 8).
The $\lambda$$\lambda$ 1180--3230 integrated flux of both the
reddening-corrected average continuum and the power-law fit is
1.94$\times$ 10$^{-10}$ erg cm$^{-2}$s$^{-1}$.

\section { The emission line spectrum }

The strongest emission features are CIV 1550 and HeII 1640 (Fig. 3,
top), while weaker emission lines are identified as OIII] 1666, NIII]
1750, etc. (see Table 2 for a list of additional emission lines in the
average spectrum, together with their tentative identification and
their reddening-corrected emission intensity).  The LW region seems
almost featureless, but a closer inspection reveals marginal evidence
of weak emission and absorption features.  The emission line near 3135
\AA~ is identified as OIII 3134, a Bowen fluorescence line excited by
the HeII Ly-$\alpha$ transition at 303.8 \AA.

\begin{figure}
\centering
\resizebox{\hsize}{!}{\includegraphics{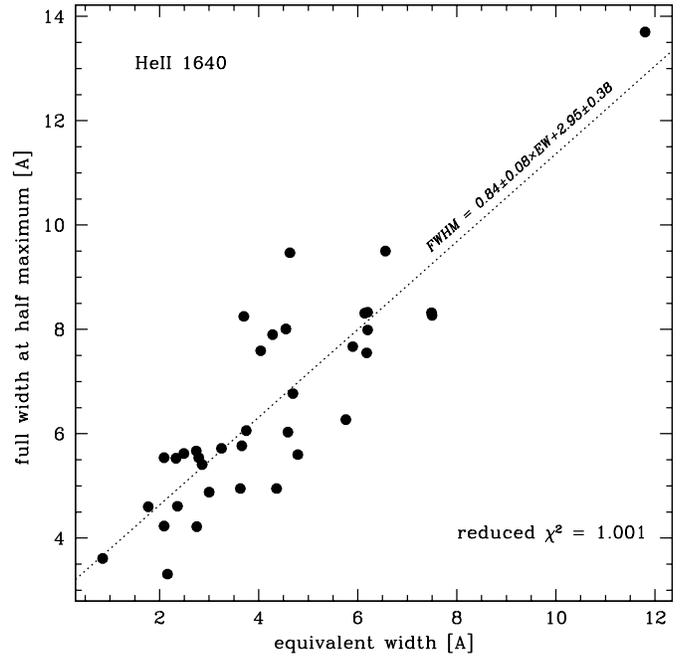}}
\caption{Correlation between the FWHM and the
equivalent width for the HeII 1640 \AA~ emission line.}.
\end{figure}

An examination of the individual spectra shows that only the emission
lines of CIV 1550 and HeII 1640 are present in all spectra, although
quite variable, while OIII] 1666 and NIII] 1750 are present in about
one half of the spectra.
The changes in intensity of the emission lines of CIV 1550 and HeII
1640 are associated to changes in the line width.  Table 3 gives the
observed equivalent width (EW) and FWHM for these emission lines,
although the limited
spectral resolution of the data has precluded very accurate
measurements.  A good correlation (rms of residuals = 1.015) exists
between the FWHM and the EW of the HeII 1640 line
(Fig. 4). The corresponding correlation for the CIV 1550 emission line
is poorer (rms of residuals = 1.579), while the correlation between
the fluxes in the HeII 1640 and CIV 1550 lines is just fair.

\begin{table}
\begin{minipage}[t]{\columnwidth}
\caption{The  intensities of the emission lines in the average spectrum
and
their proposed  identifications. }
\centering
\renewcommand{\footnoterule}{}
{\scriptsize
\begin{tabular}{ccl}
\hline
Wavelength\footnote{The listed wavelength is the mean
multiplet wavelength.} & Intensity  & 
Identification\footnote{Identifications marked with a (?) are
doubtful.}\\
(\AA) & (10$^{-13}$ erg cm$^{-2}$ s$^{-1}$) & \\
&&\\

1177  &          1.98  &             CIII(4) 75.70  \\
1291  &          1.02  &             CI(4.01) 87.82 ? \\  
1354  &          0.50  &             OI(1) 55.60 + CI(43) 54.28 +   \\
 &&CI(42) 55.84\\
1433  &          0.62  &             unidentified (also in HR Del)      
\\   
1502  &          0.73  &             SiIII 1501 (diel.) ?\\
1510  &          0.41  &             OV(64)  06.76 ?\\
1520  &          0.37  &             unidentified \\
1550  &          8.35  &             CIV(1) 49.50\\
1565  &          0.62  &             CI(UV3) 60.86 ?\\
1578  &          0.72  &             CIII(UV12.03) 77.25 ?\\
1586  &          0.93  &             OIII(19.23)  87.53 ?\\
1620  &          0.72  &             (CIII(11.72)  20.30 +\\
&& NV(53) 19.74) ?\\
1640  &          6.32  &             HeII(12) 1640.39\\
1667  &          1.26  &             OIII(0.01) 64.15\\
1702  &          0.22  &             OV(67)  07.99 ?\\
1738  &          0.30  &             NII(13.25) 42.25 ?\\
1750  &          2.32  &             NIII(0.01) 50.46\\ 
1872  &          0.63  &             unidentified   \\
1905  &          0.35  &             CIII(0.01) 06.67\\
1920  &          0.16  &             CIII(12.02) 23.13 ?\\
2308  &          1.22  &             HeII(Pa) 06.20 + ?\\
2392  &          0.37  &             HeII(Pa) 85.40 ?\\
2516  &          0.74  &             HeII(Pa) 11.21 ? + ?\\
2733  &          0.41  &             HeII(Pa) 33.32\\
3134  &          1.84  &             OIII(12) 32.79 (Bowen)\\
\hline
\end{tabular}
}
\end{minipage}
\end{table}

\begin{table}
\caption{The equivalent widths (EW) and 
FWHM for the emission lines of CIV 1550 and HeII 1640 in individual
spectra.}
\centering
{\scriptsize
\begin{tabular}{ccccc}

& \multicolumn{2}{c}{CIV 1550} &\multicolumn{2}{c}{HeII 1640} \\

SWP &  FWHM &  EW & FWHM &  EW  \\
& [\AA] & [\AA] & [\AA] & [\AA] \\
&&&&\\
08973  &      11.6 & 6.7 & 6.8 & 4.7\\
29318  &      3.7 & 6.4 & 6.3 & 5.8\\
32218  &      6.2 & 4.4 & 5.6 & 4.8\\
32899  &      6.0 & 5.5 & 8.3 & 6.1\\
33034  &      6.4 & 11.5 & 6.0 & 4.6\\
34696  &      5.1 & 4.9 & 5.0 & 4.4\\
37536  &      5.0 & 4.1 & 7.6 & 4.0\\
43442  &      7.8 & 5.6 & 4.2 & 2.1\\
44182  &      6.3 & 4.5 & 4.9 & 3.0\\
44948  &      7.4 & 5.5 & 3.3 & 2.2\\
46605  &      5.8 & 3.6 & 13.7 & 11.8\\
47057  &      7.0  & 4.5 & 4.9 & 3.6\\
47323  &      6.4 & 3.9 & 4.2 & 2.8\\
47328  &      6.7 & 4.8 & 8.2 & 3.7\\
47332  &      5.4 & 4.1 & 8.3 & 7.5\\
49365  &      8.6 & 6.9 & 9.5 & 6.6\\
49366  &      7.8 & 5.5 & 7.9 & 4.3\\
50099  &      6.5 & 4.0 & 5.5 & 2.3\\
50100  &      5. & 3.6 & 8.3 & 7.5\\
50596  &      6.3 & 4.3 & 7.7 & 5.9\\
52886  &      4.0 & 4.4 & 5.5 & 2.8\\
52887  &      4.5 & 4.4 & 5.4 & 2.9\\
53809  &      6.1 & 4.8 & 8.3 & 6.2\\
53810  &      7.3 & 5.8 & 5.7 & 2.7\\
54590  &      10.2 & 4.6 & 9.5 & 4.6\\
56240  &      8.7 & 6.6 & 4.6 & 1.8\\
57030  &      6.3 & 6.6 & 5.7 & 3.2\\
57031  &      5.9 & 5.2 & 8.0 & 4.6\\
57032  &      6.1 & 5.0 & 5.8 & 3.7\\
57033  &      6.0 & 6.1 & 6.1 & 3.7\\
57034  &      6.7 & 6.9 & 5.6 & 2.5\\
57035  &      7.1 & 8.4 & 7.6 & 6.2\\
57039  &      5.6 & 4.8 & 3.6 & 0.9\\
57042  &      4.0 & 4.1 & 5.5 & 2.1\\
57047  &      3.1 & 2.0 & 8.0 & 6.2\\
57055  &      5.7 & 4.4 & 4.6 & 2.4\\

\end{tabular}
}
\end{table}

\begin{table}
\begin{minipage}[t]{\columnwidth}
\caption{Emission lines present only in a few spectra  and {\sl not}
listed in Table 2.  }
\centering
\renewcommand{\footnoterule}{}
{\scriptsize
\begin{tabular}{cll}
\hline
Wavelength\footnote{The listed wavelength is the mean
multiplet wavelength.} & Identification\footnote{Most identifications
are tentative.} &  SWP spectrum \\
&&\\
1190 &          SIII(1) 90.20  &  29318, 37536, 44182, \\
     &                         & 53810, 56240\\             
1243 &          NV(1) 40.15  & 49365\\
1250 &          CIII(9) 47.38   & 57039\\
1285 &          CI(5) 80.62  & 56240\\
1371 &          OV(7) 71.29          & 08973\\
1412 &          NI(10) 11.97   &  43442\\
1474 &          SI(3,4) 73.77  &  50596\\
1713 &          SiII(10) 11.0   &  57034\\
1727 &          ?        &  53810 \\
1775 &          ?            &  57039\\
1782 &          FeII(191) 85.27 & 29318,\\ 
1890 &          SiIII(1) 92.03  &  50099\\
1910 &          CIII(0.01) 08.73   &  08973\\
\hline
\end{tabular}
}
\end{minipage}
\end{table}

\subsection {The emission lines present in individual spectra only }

One peculiar aspect of the line spectrum of T~Pyx is  that the emission
lines show substantial changes in 
intensity and detectability in
the individual spectra, in sharp contrast to the near constancy of
the continuum  (See the SWP spectrum gallery in Fig.
5).

\begin{figure}
\centering
\resizebox{\hsize}{!}{\includegraphics{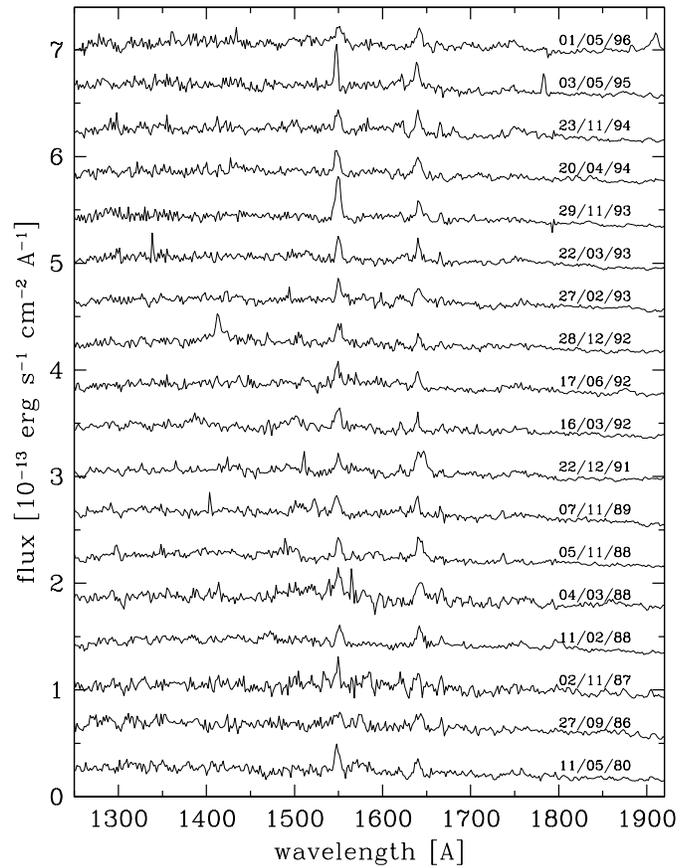}}
\caption{SWP observed spectra gallery. Spectra are offset by 4
10$^{-14}$ erg s$^{-1}$ cm$^{-2}$ A$^{-1}$. The near constancy of the
continuum contrasts with the changes in the emission lines. }
\end{figure}

This behavior may cast some doubt on the reality of some emission
features, as some of them may be spurious features produced e.g. by
cosmic rays hits, due to the long exposure times (more than three
hours).  A detailed and specific inspection of all the line-by-line
(SILO) spectral images has  confirmed, however, that several of these
features are real emission lines, with only a few being spurious. It
is surprising, and somewhat worrisome, that in the extracted spectra
of LWP12791 and LWP22052 strong and wide emissions features are
present in the region 3100-3250 \AA~, but absent in the SILO
image. All SILO images show that the strong Ly$\alpha$ emission is of
geocoronal origin, as expected on account of the long exposure times.

Table 4 gives the wavelength and the (tentative) identification for
the emission feature considered as real.  For many of these emissions
the identifications are rather uncertain since they are uncommon in
other CVs.

\subsection {The features outside  the  spectrum}

It is generally assumed that the shape of the large entrance aperture
of the IUE SWP spectrograph is that of a rectangle whose projected
size on the camera faceplate has approximate dimensions of
10$\times$20 arcsec. The direction of the dispersion is nearly
perpendicular to the major axis of the aperture. Therefore, on the
SILO images of an extended object (nebula), one expects to see
spectral features on both sides of the star spectrum, out to a distance
 of $\pm$ 10 arcsec.

Actually, the IUE large apertures have parallel sides and rounded ends
(Bohlin et al. 1980). The full length to the tips of the rounded ends
and the width of the SWP's large aperture were measured before flight
and the effective dimensions are  23.0 $\times$ 10.3 arcsec.  The
IUE NEWSIPS manual, Table 2.1 (Garhart et al. 1997)  instead gives a
large aperture length of 21.65 $\pm$ 0.39 arcsec and a large aperture
width of 9.07 $\pm$ 0.11 arcsec, but the reported large aperture area
(215.33 arcsec$^2$) is quite a bit larger than the product of the quoted
length and width, and it is  probable  that  the  length does not
include
the two rounded ends. The precise form and size of the IUE large
aperture is apparently not well-defined (see also Gonzalez-Delgado and
Perez 2001), but can be considered as intermediate between a rectangle
and an ellipse. In the present study we  conservatively chose to
consider only features at a distance of $\le$ 10 arcsec on either side
of the central strip for well-centered objects. The only notable
exception is  SWP47332 (see below).

\begin{figure*}
\centering
\includegraphics[width=18cm]{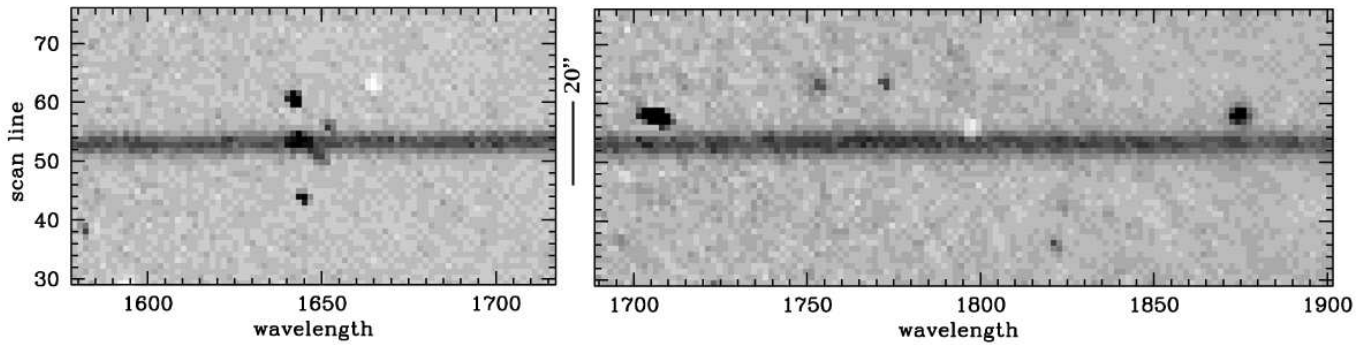}
\caption{Line-by-line images of SWP47332 (left) and SWP 57055 (right).
The two symmetrical knots on either side of the HeII 1640 line in the
left image fall very near the border of the large aperture. Their
reality is uncertain, see the text for details. The bright knots just
above the spectrum in the image on the right are those responsible for
the emission lines at 1703 and 1873 \AA~ in Fig. 1.}
\end{figure*}

The inspection of all SILO images led to the detection of several
emission spikes {\it outside} the central strip of the spectrum but
still inside the 20 arcsec large aperture. Some pseudo-emission
features are produced by the {\it permanent} presence of ``hot pixels"
(for example, a typical one falls near 1750 \AA, nearly the same
wavelength as a well-known NIII] line, though outside the typical
position of well-centered spectra).  It is also possible that several
of these spikes are caused by cosmic ray hits mimicking the PSF of
{\sl IUE}.  However, there are cases in which the spikes either are
strictly associated with known emission features present in the (main)
spectrum or show a common pattern hardly attributable to the impact
of cosmic rays.

For example: In SWP49365 a spike at 1243 \AA~ is at the same
wavelength as a spectral feature,  probably NV 1240; in SWP43442 a
knot
at 1414 is at nearly the same wavelength as a spectral feature at 1412
\AA; in SWP33034, SWP53809, and SWP47057, there are spikes near 1545
\AA~  that seem associated with CIV 1550; in SWP57055 there are
two
strong spikes at 1703 and 1873 \AA, both lying on the same side of the
spectrum, at the {\it same} distance (about 9 arcsec) from it (see
Fig.  6 right; as already mentioned in Section 2 they are present as
strong
emission lines in the INES extracted spectrum and flagged as ``cosmic
rays"); the same in SWP54590 at 1751 and 1934 \AA; finally in SWP08973
a spike at 1909 \AA~ is associated with an emission feature at 1910
\AA~ on the spectrum (probably CIII] 1909).

The case of SWP47332 is instead quite problematic because  there
are apparently four spikes near 1640 \AA~ (Fig. 6 left) and also four
ones near
1400 \AA~ (not in the same pattern, although one is  the same
distance from the spectrum at both wavelengths). The two stronger
spikes on both sides of HeII 1640 would fall outside the large
aperture if one assumes the ``standard" 20 arcsec size, but they would
fall just near the border if one adopts the 23.0 arcsec size given by
Bohlin et al.  (1980). A comparison with the spatial size
(perpendicular to the dispersion ) of the geocoronal Ly-alpha (that
fills the large aperture) shows that the separation between the two
external features near $\lambda$ 1640 is close to the spatial width of
the geocoronal Ly-alpha, although the saturation effects in the
geocoronal line, together with the difference in the spatial point
spread function for the two wavelength regions (Cassatella et al.,
1985, Garhart et al., 1997), makes it difficult to make a definite
statement about the true nature of the two external features. This
particular example highlights the difficulties one encounters in
establishing the reality of the ``external" features in IUE spectra,
especially when nature also maliciously conspires to place two
symmetrical knots on either side of a well-known spectral line.

In general, if the external spikes correspond to spectral features,
they may originate in the extended region (nebula) surrounding the
system, although  that only one or, at most, a few external
features (and not an entire spectrum of several emission lines) are
present in an image is of difficult to explain physically.  On the
other hand, an examination  of the SILOs of other
faint emission line objects (Seyfert Galaxies), made for comparison,
 has not revealed as many spikes features as in T~Pyx, although the
spectra
were taken with similar exposure times.

An analysis of the distribution of spikes inside and outside the large
aperture has not been conclusive in demonstrating that the
distribution inside is statistically larger than that outside.  Apart
from small numbers, the  probable reason for this is that the
background
varies substantially in different regions of the image, and even small
variations in the threshold (in terms of $\sigma_{\rm bgd}$) for
automatic spike detection give different results. If one only uses
``bright" spikes, the number within the aperture is often larger than
that expected from the average in the frame; but this is difficult to
quantify in a totally objective way since the ``brightness" of the
spikes also depends  on the position in the frame. Even so, the result
for bright spikes, together with the comparison with similarly exposed
spectra, would appear a strong argument in favor of at least some of
the spikes being real line emissions.

There are. however, no obvious, known candidates that would produce
spikes near the edge of the aperture. Figure 7 shows the {\sl IUE}
aperture during the observations in Figure 6 superimposed on the HST
image of T~Pyx and its surroundings (Shara et al, 1997). The aperture
is plotted with a standard $10''\times 20''$ size, although the long
axis is  probably 15\% longer, as explained above. The lack of
prominent knots in the shell near the rounded edge of the aperture is
not in itself sufficient for excluding the reality of the spikes: the
HST
image is a composite of observations taken in 1994-95, while the {\sl
IUE} observations were taken in 1993 and 1996 and we know that the
spikes are short-term transients, e.g. the ones in SWP 57055 were not
present a day before in SWP 57047 (although this point may be
considered an argument against the interpretation that they are
emission lines). The absence of candidates certainly weakens the
interpretation that the spikes are emission lines rather than cosmic
rays in the case of SWP 47332. There are instead enough knots at
5--6$''$ from T~Pyx that an origin in the nebula for the spikes in SWP
57055 cannot be ruled out.

\begin{figure}
\centering
\resizebox{\hsize}{!}{\includegraphics{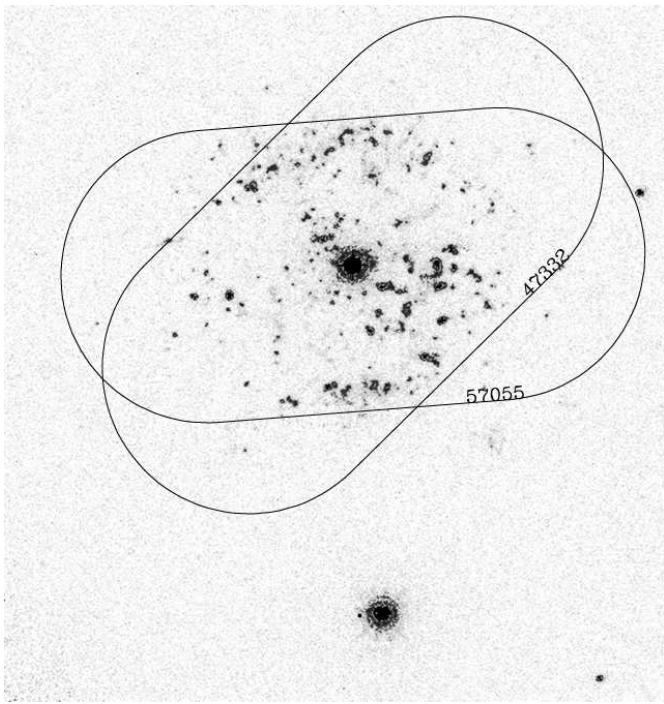}}
\caption{{\sl IUE} slit position during the observations reported in
Fig. 6 superimposed on an HST image of T~Pyx and its ``shell''. The
{\sl IUE} aperture is drawn with a size of $10''\times 20''$ (tip to
tip), although the long axis is probably 15\% longer. No obviously
bright candidates for the transient features detected in 1993 and 1996
by {\sl IUE} are evident near the edge of the aperture in the
composite HST image taken in 1994-1995.}
\end{figure}

\section {The absorption lines and the model spectra }

In spite of the limited spectral resolution, the good S/N in the
average spectrum has allowed  positive detection of several
absorption features (Fig. 3, top).  They are all listed in Table 5,
together with their corresponding equivalent widths and the
identifications.  Most of them are identified as well-known resonance
lines of once ionized species.

An interstellar contribution to the zero-volt component in some of
these lines is  expected, but the presence of intrinsic variations
in
individual spectra (especially for the SiII 1260 line) and the
weakness of the absorption near MgII 2800 indicate that it is not
dominant.  Unfortunately, the limited resolution of the {\sl IUE} low
resolution spectra has not allowed  the intrinsic profile
of the lines to be appreciated and  the probable presence of more
than one
contributor to be detected.

\begin{table}
\caption{Equivalent widths and  identifications for the absorption lines
in the average spectrum.}
\centering
{\scriptsize
\begin{tabular}{ccl}

Wavelength & EW [\AA] &      Identification \\
&&\\
     1191    &    0.89 &    SiII(5) 90.41 + SIII(1) 90.20   \\
     1259    &    1.15 &    SiII(4) 60.42 + FeII(9) 60.54\\ 
                        && + SII(1) 59.52\\
     1303    &    1.01 &    OI(2) 02.17 + SiII(3) 04.37 \\
     1334    &    0.73 &    CII(1) 35.10\\
     1371    &    0.40 &    NiII(8) 70.20\\
     1529    &    0.52 &    SiII(2) 26.71\\
     1628    &    0.49 &        ? \\
     1727    &    0.82 &        ? \\
     1807    &    0.50 &    SiII(1)  08.01\\
     1816    &    0.14 &    SiII(1)   16.92\\
     2376    &    1.35 &    FeII(2) 73.73    \\
     2475    &    1.00 &    TiII(2) 78.64 \\
     2600    &    0.75 &    Fe II(1) 99.39  + MnII(1) 05.70 \\
     2799    &    1.06 &    MgII(1) 97.92 + \\
     2853    &    0.41 &    MgI(1) 52.12      \\

\end{tabular}
}

\end{table}

It is probable that most of the absorption lines originate in the
accretion disk itself.  Wade and Hubeny (1998) present a large
grid of computed spectra from steady-state accretion disks in luminous
CVs.  Disk spectra corresponding to twenty-six different combinations
of accretion rates and WD masses are computed and tabulated for six
different disk inclination angles $i$.  The wavelength coverage of the
models ranges from 800 to 2000 \AA~, so it is possible to
compare them with the {\sl IUE} spectra taken with the SWP camera
(1160-1960 \AA).  The spectral features we  used for the fitting
are the intensity and width of the undisplaced absorption lines (e.g.
SiII 1260, OI + SiII 1305, CII 1335, etc).  Most of the 156 different
models were compared with the reddening-corrected average SWP
spectrum of T~Pyx but, disappointingly, while there is quite a range
of values that gives a satisfactory agreement with the observed depth
and shape of the absorption lines, no single solution is convincingly
valid at the same time both for the continuum distribution and the
depth and shape of the absorption lines.  The best fits for the shape
and depth of the lines (models $ee$ and $jj$ at rather low
inclination angles) correspond to high values for the white-dwarf mass
($M_1$ $\sim$1.21$M_{\odot}$ and $\sim$1.03$M_{\odot}$), but give a
continuum that is too steep.

\section {Discussion}

In the following three subsections  we will first compare 
the UV spectrum of T Pyx with that of other ex-novae, then we will
suggests some mechanisms  to explain the origin of the UV emission
lines and their variations,  and finally we will consider the origin of 
the observed optical and IR magnitudes of T Pyx. 

\subsection {T~Pyx and other ex-novae in the UV}

We  compared the dereddened  UV spectrum of T~Pyx with 
all the available  IUE  spectra of  novae  in
quiescence  (about 18 objects, including   V603 Aql, RR Pic, HR Del,
V533 Her,  V841 Oph, etc.).

Surprisingly, the best agreement for the shape of the continuum and
the intensity of the emission lines has been found with the very-slow
nova HR Del (see also Selvelli and Friedjung, 2003), whose UV spectrum
would be almost indistinguishable from that of T~Pyx, were it not for
the two absorption components in the NV 1240 and CIV 1550 resonance
lines that are prominent features in HR Del (Fig. 8).
This remarkable similarity in the continua of the two stars is quite a
challenge to common interpretations: T~Pyx is allegedly observed at
rather low inclination ($i\sim$ 15-20 degrees) and its WD must be very
massive (about 1.2-1.4 $M_{\odot}$) in order to guarantee outbursts
with a very short recurrence time (see also Paper II).  HR Del,
instead, is observed at medium inclination ($i\sim$ 45 degrees) and
its WD is allegedly on the light side (about 0.6 $M_{\odot}$) (see
Selvelli and Friedjung, 2003 for a discussion on these latter
parameters).  If we anticipate from Paper II that the accretion rates
in the two stars are quite similar, then one would expect  a
much hotter (steeper) continuum for T~Pyx.
The fact that T~Pyx and HR Del show the same intensity in the
HeII 1640 emission line suggests that these (in principle very
different) objects also have very similar temperature and continuum
distribution  in the EUV region, since the 1640 \AA~ line is a
recombination line of He$^{++}$, whose ionization is controlled by the
radiation field shortward of 228 \AA.

Another peculiarity of T~Pyx is its absorption spectrum.  Blue-shifted
absorptions components in the resonance lines of high ionization
species (CIV, SiIV, NV) are generally seen in the UV spectra of
luminous (high mass-transfer rate) cataclysmic variables observed at
low inclination (e.g. Nova-like, dwarf novae in outburst, the old
novae HR Del and V 603 Aql).  The absence of any such signature of
outflow in T~Pyx (which is luminous and observed at low inclination)
indicates that at least one basic ingredient for the onset of outflow
is missing in this system.

Moreover, the absence in most spectra of T~Pyx of both the NV 1240 and
the SiIV 1398 resonance lines is also surprising, since these emissions
are common in most UV spectra of cataclysmic variables, and, usually,
because of their different sensitivity to temperature effects, the
absence of one ion is complemented by the presence of the other (while
the intermediate temperature ion CIV 1550 always tends to be
present).  We can tentatively interpret the absence of SiIV as due to
a very hot boundary layer whose hard radiation keeps silicon in
ionization states higher than Si$^{+3}$, but in this case it is hard
to understand why there is no evidence of the NV 1240 lines either in
emission or in absorption.

In the old nova V603 Aql, which is observed at almost the same (low)
inclination as T~Pyx, both NV 1240 (in absorption) and SiIV 1400 (in
emission) are present.  Also in the old nova RR Pic the line
excitation is much higher than in T~Pyx: the NV 1240 \AA, CIV 1550
\AA, and HeII 1640 \AA~ emission lines are much stronger than in T~Pyx.
This point will be further discussed in Paper II in the context of the
alleged SSS nature of T~Pyx, as proposed by Patterson et al. (1998).

Common nebular lines (e.g.  NIV] 1483, SiIII] 1892, CIII] 1909, etc) are
usually absent in the UV spectrum of T~Pyx, in contrast to the
presence of a (composite) nebular shell surrounding the central
object.  It is tempting to explain this with the fact that  the nebula 
has a quite small covering factor because of
its knot structure and
therefore most of the Lyman continuum radiation of the central hot
source leaves the star undisturbed and is not degraded into line
radiation (HI emissions plus coolants).  However, the presence of
conspicuous emission lines of HI and HeII casts some doubts on this
interpretation.  Another possibility is that the near absence of
nebular lines could simply be a density effect, the knots'
condensations having electron densities higher than the critical ones
for the relevant emission lines.

\subsection {Speculations on the origin of the UV emission lines 
and their variations}

The emission line spectrum of T~Pyx shows some peculiarities whose
interpretation is not straightforward:

1.  The strongest emission lines (CIV 1550, HeII 1640) are variable in
intensity and width.

2.  Some lines are observed in a few spectra only and are uncommon in
CVs and/or lack reliable identification.

3. Some lines appear inside the slit (not on the spectrum) at uncommon
wavelengths and varying positions.

The observed emission-line variations, which in some cases take place
on time scales as short as a few hours (see the spectra taken in
sequence in May 1996), are not caused by instrumental effects, i.e.
different centerings in the {\sl IUE} aperture that could 
produce different contributions from the nebular shell.  In all the SILO
images we have examined (with the few exceptions mentioned in \S 4.2),
there was no evidence of any contribution from an extended region.

The {\sl IUE} Roll angle (which determines the slit's position angle)
is fixed during one exposure, but varies at different epochs.
Therefore, knots that may be in the $10\times 20$ arcsec aperture at
one epoch may not be at another, although knots within 10 arcsec of
the star are always in the aperture.  Even for these, at different
position angles the same knot will produce lines that are not only at
a different distance from the main spectrum (or even {\it on} it) but
also at different wavelengths from epoch to epoch (up to $\pm 8$ \AA~
depending on the geometry inside the slit). This may help explain the
sporadic appearance of lines on and outside the spectrum but, of
course, not the variations of the known lines.

Also, the variations can hardly be ascribed to aspect-dependent and/or
orbital and geometrical effects (bulges, hot spots, etc.) because the
system is allegedly viewed nearly pole-on ($i \leq 20 \degr$, Paper II)
and  the average exposure time of all spectra is longer than the
orbital period  so that orbit-dependent effects should be averaged over
one orbital period or more.
The observed variations in the lines of the spectrum must take place
in a region whose angular size is on the order (or less) of the PSF of
{\sl IUE} ($\leq $ 2.7 $\arcsec$).  This region contains the binary
system
and the innermost nebula (and that part of the nebula that is
projected on the compact source).

\begin{figure}
\centering
\resizebox{\hsize}{!}{\includegraphics[angle=-90]{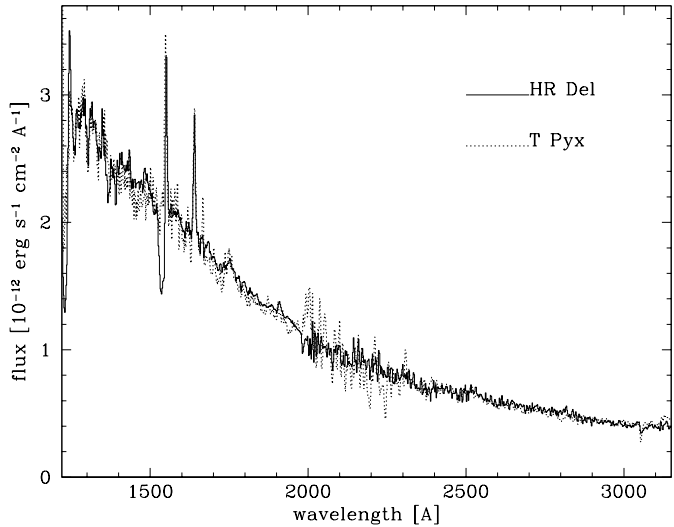}}
\caption{Comparison between the spectra of T~Pyx and HR Del. The two
spectra are almost indistinguishable except for the   P Cyg component in
the  CIV  1550  line of HR Del. The spectrum of T~Pyx has been
scaled to the flux scale of HR Del.  }
\end{figure}

In other CVs, intrinsic variations in the emission lines are observed
in systems that are not stable accretors (e.g., dwarf novae from low
to high states, etc.) when they suffer $\dot{M}$ changes in an
optically thin accretion regime.  This is apparently not the case for
T~Pyx.  The high luminosity of the UV continuum, remarkably constant
over 16
years, indicates stable accretion in an optically thick regime.

Besides the difficulty of explaining the variations, the problem also
remains  with the unidentified or poorly identified features.  It
is  probable that these emission lines may come from regions other
than
the hot component (accretion disk plus boundary layer), so we
tentatively suggest these non mutually-exclusive regions and/or
mechanisms:

{\it 1) The (irradiated) chromosphere of the M dwarf companion.}

Several short-period detached binary systems are known - e.g.: V471
Tau (Guinan 1990, Young et al., 1988), BE UMa (Ferguson et al., 1991),
Feige 24 (Vennes et al. 1991) -  which show irradiation induced effects:
the atmosphere of the companion star, usually a low-MS star, is heated
by the flux of the WD leading to an enhancement of line emission.
These systems are thought to be pre-cataclysmic binaries (de Kool and
Ritter 1993); see also Jomaron et al. (1993).  In some of these
systems, e.g.  V471 Tau, the atmosphere of the K star is inhomogeneous
and there is evidence of external plasma around it (Young et al. 1991,
Guinan 1990).  These atmospheric structures may be related to the
cool loops or active region plumes observed in the Sun, but are much
more extensive in size and are unstable, with a lifetime of a few
days.

We suggest that a similar class of phenomena could also be present in
T~Pyx, where transient jets or loops from the cool star, excited by
the strong UV radiation of the hot component, may be responsible for
the variations and peculiarities of its spectrum.  However, while the
above-mentioned objects are semi-detached binaries in which the
luminosity and temperature of the WD is not particularly high, T~Pyx
is instead a short-period semi-detached system with high ultraviolet
luminosity
($L_{\rm UV} \sim 100 L\odot$, Gilmozzi et al. 1998, and Paper II) and
rather high average $T$ ($\sim$34100 K), so that much stronger
induced effects are expected.  Urban (1988) has already pointed out the
importance of irradiation effects on the spectrum of T~Pyx.

{\sl \it 2) The enhanced activity in the secondary and magnetism.}

It is well known that most M dwarfs are characterized by a surface
magnetic activity that is manifested by non thermal emission in a hot
outer atmosphere (see Hawley 1993 for a review).  The magnetic
activity increases monotonically toward later spectral types, until at
type M5 and later all dwarfs are dMe or active and show
high-temperature UV emission lines.  Rotation also plays an important
role in this class of phenomena, therefore, it is  possible that
binarity with a close companion might increase these effects by
tidally-induced rapid rotation, together with irradiation and heating.
Flare activity is associated with phenomena like coronal loops and
coronal mass ejection that have sizes on the order of the stellar
radius and time scales on the order of hours (see the observations of
AU Mic with EUVE by Katsova, 1996).  Also, if the WD in T Pyx is
mildly magnetic, the two stars are  probably interconnected by a
magnetic field that acts as a guide line for the material streaming
from the secondary, and the interaction of the two magnetic fields may
heat the plasma between the two stars.
A representative spectrum of an active star (AD Leo) is given in Fig. 3
(bottom); the three strongest emission lines are those of CIV
$\lambda$ 1550, HeII $\lambda$ 1640, and OIII $\lambda$ 1660, with
relative intensities similar to those observed in T~Pyx.

In the context of these two models, the absence in the UV spectrum of T
Pyx of the MgII 2800 emission line, a well-known indicator of
chromospheric activity, is quite disturbing.  We can only invoke the
somewhat {\it ad hoc} explanation that the MgII emission is suppressed
by the corresponding absorption component of the accretion disk and by
the interstellar medium.

{\sl \it 3) Illumination of (moving) knots of the nebula in front of the
compact object.}

Shara et al. (1997) report that a few knots are observed to fade or
brighten significantly on a timescale of months and interpreted this
in terms of collisions and shocks between successive generations of
ejecta.  From the comparison of the size of the nebula at different
epochs, they derived an upper limit of about only 40 km s$^{-1}$ (but
 assumed d = 1500 pc) on the systematic expansion velocity of the
knots.  Instead, both Margon and Deutsch (1998) and O'Brien and Cohen
(1998), in a study of the alleged jet components in the H$\alpha$ line
(Shabaz et al 1997), identified these features as due to the [NII]
6548, 6584 lines and interpreted their wavelength shifts as produced
by an expansion velocity  of about 500 km s$^{-1}$ along the line of
sight 
through the center of
the shell. If some of these ``variable knots" happen to lie within 
the {\sl IUE}
PSF they may also explain the emission line changes and the unusual
spectroscopic signatures in the UV spectrum of T~Pyx.

To summarize this rather speculative section, we propose that the
peculiar spectrum of T~Pyx may be due  to complex velocity
fields in the material (cool loops, active region plumes, transient
jets, coronal loops, etc) that streams from the active companion
toward the WD and/or to the transient appearance of high-velocity
emitting knots in front of the compact object.
That in individual UV spectra only a few
extra emission lines (most of them of dubious identification) are
observed, instead of a spectrum, means that interpretation is  difficult 
and
that  quite uncommon physical conditions  are required.

The data we have are too sparse to test other possible effects
contributing to the unusual behavior of T~Pyx.  For example, we have
considered precession in the system (partially) hiding and revealing
at different epochs a possible hot spot as one explanation for the
variability in the emission lines, but much more intensive
observations would be necessary to test this hypothesis.

We  also considered other more ``exotic" scenarios in view of the
peculiarities of T~Pyx, for instance, that some of the uncommon,
sporadic lines may be typical lines that are blue- or red-shifted in a
jet.  But apart from the occasional, correct wavelength ratio between
lines being similar to that of HeII/CIV or NIII]/HeII (at a few to
several thousand km s$^{-1}$), there is little evidence to support this
hypothesis.

\subsection { The origin of the optical and IR magnitudes}

The extrapolation to the V band of the power-law fit of the UV
continuum {/bf $F_{\lambda}=4.28\ 10^{-6}\ \lambda^{-2.331}$ erg
cm$^{-2}$s$^{-1}$ A$^{-1}$} yields a flux $F_{5500} = 8.17 \times
10^{-15}$ erg cm$^{-2}$s$^{-1}$ A$^{-1}$ or $m_{\rm v}\sim$ 14.10.
After proper reddening with A$_{\rm v}$=0.78, the predicted V
magnitude is 14.88.  Similar calculations for the B band yield a flux
= $1.34\times 10^{-14}$ erg cm$^{-2}$s$^{-1}$ A$^{-1}$ or $m_{\rm
b}\sim$ 14.23.  After reddening using A$_{\rm B}$=4.1$\times
E_{B-V}$=1.03, this becomes $B=15.26$.  Therefore, the observed
$B=15.5$ (Schaefer 2005) and $V=15.3$ (Downes at al.  1997, Szkody
and Feinswog 1988) magnitudes of T~Pyx fall quite close to the values
defined by the tail of the power-law fit to the UV continuum.  This
indicates that $m_{\rm B}$ and $m_{\rm V}$ are dominated by the hot
component (accretion disk). That the extrapolated $B-V$ value
(after reddening) is close to 0.38 and larger than the observed one
(close to 0.20) is due to the fact that  the
spectrum departs from the power-law extrapolation at longer wavelengths
(see below and
Fig. 9).

\begin{figure}
\centering
\resizebox{\hsize}{!}{\includegraphics[angle=-90]{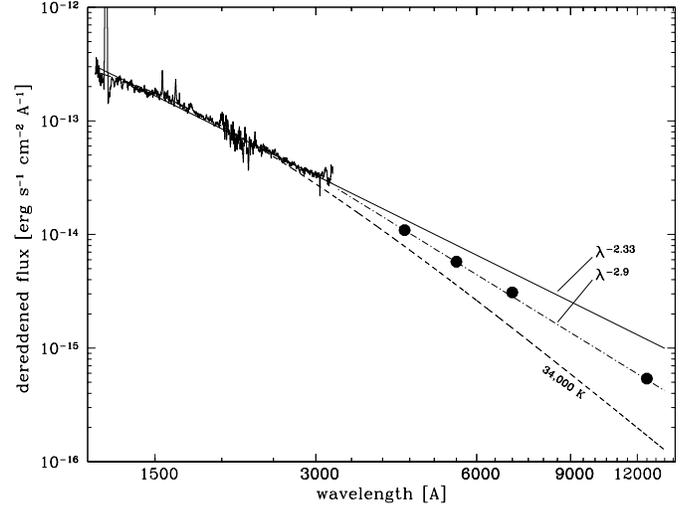}}
\caption{UV-Opt-IR ``spectrum" of T~Pyx. The B, V, R, J average fluxes
are indicated. The hot source is clearly the main contributor at all
wavelengths (the departure from the $-$2.33 power-law at long
wavelengths probably indicates that these fluxes arise in the outer
part of the disk, where it becomes optically thin or reaches a
physical edge). The $\alpha$=-2.9 power-law is an aid to the eye.  The
black body that best ``fits" the UV continuum (though meaningless for
an accretion disk) is also plotted. }
\end{figure}

The extrapolation to 12500 \AA~ (J band) gives 1.20 10$^{-15}$ erg
cm$^{-2}$s$^{-1}$ A$^{-1}$.  After conversion to J magnitudes and
reddening with $A_{\rm J}$=0.87$\times$$ E_{B-V}=0.22$, we obtain
J=13.52, to be compared with the observed J=14.4-14.65 (Szkody and
Feinswog 1988; Weight 1994) Therefore also in the J band, it is the IR
tail of the hot source that is the main contributor to the observed
magnitude.

This progressive departure from the power-law distribution is most
probably an indication that the largest contribution of the disk to
these wavelengths comes from its outer region, where it becomes
optically thin (or reaches a physical edge). With a full spectral
coverage, it should be possible to determine the temperature of this
region. Judging from the optical and infrared colors, this should not
be far from 10,000 K.

Szkody and Feinswog (1988) found a sinusoidal modulation in the J
light curve of T~Pyx from J=14.34 to J=14.50 with P = 100 min (single
sine solution) or P = 200 min (double sine solution), which was
interpreted as due to ellipsoidal variations in the secondary, that,
allegedly, provided a substantial contribution to the IR light.  They
then applied the technique of ellipsoidal fitting to derive the system
inclination (close to 90 degrees) and the mass ratio q (close to 4).

This interpretation, however, is compatible with the observations only
if the distance to T~Pyx is quite low (and they actually suggested a
value close to 350 parsecs).  In fact, from various relations
available in the literature between $M_J$ and other stellar parameters
such as the mass, the spectral type, etc.  (Henry and Mc Carthy 1990,
Henry and Mc Carthy 1993) or from the observed J values for objects in
clusters with known distances (Legget and Hawkins,1989), it follows
that the M4-M5 V  secondary star in T~Pyx (see the Introduction) has an
absolute magnitude $M_J
\sim$ 7.8 (or dimmer).  This value, together with the observed $J \sim
14.5$, and A$_J$=0.22 would imply a distance to T~Pyx of only 198 pc
(or smaller), not compatible with any value reported in the literature
whose lower limit is about 1050 pc (Shara, 1997), while a value near
3500 pc is the most plausible (see Paper II).  If the largest value
is
correct, the
apparent J magnitude of the secondary star would be about 20.7, and
would thus
contribute to less than 1 percent of the observed J flux.

Therefore the IR tail of the hot source is the main
contributor to the observed J=14.4 magnitude.  It is possible,
however, that heating (irradiation) of the secondary by the accretion
disk (see \S 6.2) could provide a partial contribution to the J light
and be responsible for its modulation.  In any case, the presence of a
definite modulation in the J curve and the absence of a similar
modulation in the optical, where several photometric periods have been
reported, are  difficult to interpret if, allegedly, the system is
seen at the inclination of $\sim$15 degrees (Warner 1995).

\section {Conclusions }

The constancy of the UV continuum energy distribution over 16 years
of {\sl IUE} observations is quite remarkable in an object that
belongs to the CVs class.  This behavior contrasts with the
observed changes both in intensity and detectability for the UV
emission lines.  The strongest emission lines (CIV 1550 and HeII 1640)
are clearly variable in intensity and width.

The origin of these variations is unclear, so we suggest that the
peculiar behavior of T~Pyx may be due either to complex velocity
fields in the material  that streams from the active companion
toward the WD (cool loops, active region plumes, transient
jets, coronal loops, etc.) and/or to the transient appearance of 
high-velocity
emitting knots from the nebular shell in front of the compact object.
    
The best single-curve fit to the dereddened ($E_{B-V}=0.25 \pm 0.02$)
UV continuum is a power-law distribution  $F_{\lambda}=4.28\ 10^{-6}\
\lambda^{-2.33}$  erg cm$^{-2}$s$^{-1}$ A$^{-1}$, with a small
uncertainty of $\pm$0.04 in the index.  The observed B=15.5, V=15.3,
and J=14.4 magnitudes of T~Pyx fall quite close to the values defined
by the tail of this power-law.  This clearly indicates that the hot
source (accretion disk) is the main contributor at all wavelengths and
that
the contribution from the secondary star is negligible, also   in the
IR.

\begin{acknowledgements}

We gratefully thank Angelo Cassatella for his participation in the
observations and in the preliminary phase of this investigation. We
also thank Rosario Gonzalez-Riestra for helpful technical discussions
in the early stages of the long gestation period of this article.  We
also thank the referee for bringing to our attention the recent paper
by Schaefer (2005) that appeared just as  this paper was
being finished.

\end{acknowledgements}


\begin{thebibliography}{}

\bibitem[Bohlin et al.(1980)]{1980A&A....85....1B} Bohlin, R.~C.,
Sparks, 
W.~M., Holm, A.~V., Savage, B.~D., \& Snijders, M.~A.~J.\ 1980, \aap,
85, 1 

\bibitem[ ]{ }Bruch A.,  Duerbeck H.W.,   Seitter W.C.,   1981,  Mitt.
Astr. Ges. , 53,34 

\bibitem[ ]{ }Bruch A.,  Engel  A., 1994, \aaps\ 104,79 

\bibitem[Cassatella et al.(1985)]{1985A&A...144..335C} Cassatella, A., 
Barbero, J., \& Benvenuti, P.\ 1985, \aap, 144, 335 

\bibitem[ ]{ }de Kool M., Ritter H.,  1993,  \aap\  267,397

\bibitem[Downes et al.(1997)]{1997PASP..109..345D} Downes, R., Webbink, 
R.~F., \& Shara, M.~M.\ 1997, \pasp, 109, 345 

\bibitem[ ]{ }Duerbeck H.W.,  Space Sci.Rev., 45,95

\bibitem[ ]{ }Duerbeck  H.W. Seitter W.C.  1979, The Messenger 17,1

\bibitem[ ]{ }Ferguson D.H., et al.  1987,  \apj\ 316, 399

\bibitem[Garhart et al.(1997)]{1997IUENN..57....1G} Garhart, M.~P.,
Smith, 
M.~A., Turnrose, B.~E., Levay, K.~L., \& Thompson, R.~W.\ 1997, IUE NASA 
Newsletter, 57, 1 
 
\bibitem[ ]{ }Gilmozzi R.,  Selvelli P., Cassatella A., 1998, in
"Ultraviolet Astrophysics Beyond the Final Archive " ESA-SP 413, p. 415 

\bibitem[Gonz{\'a}lez Delgado \& P{\'e}rez(2000)]{2000MNRAS.317...64G} 
Gonz{\'a}lez Delgado, R.~M., \& P{\'e}rez, E.\ 2000, \mnras, 317, 64 

\bibitem[ ]{ }Gonzalez-Riestra,  R., Cassatella, A., Wamsteker W., 2001,
\aap\ 373,730

\bibitem[ ]{ }Guinan E.F., 1990, in "Evolution in Astrophysics" ESA-SP
310,   p. 73

\bibitem[ ]{ }Hawley S.L., 1993, \pasp\ 105,955  

\bibitem[ ]{ }Henry T.J.,  Mc Carthy D.W., 1990  \apj\ 350,334,

\bibitem[ ]{ }Henry T.J.,   Mc Carthy D.W., 1993 \aj\ 106,773 

\bibitem[ ]{ }Jomaron C.M., et al. 1993,  \mnras\  264,219

\bibitem[ ]{ }Katsova M.M., Drake J.J., Livshits M.A.,  1996 in
"Astrophysics in the EUV", IAU Coll. 152, S. Bowyer R.F. Malina (eds.)
Kluwer  Acad. Bubl., p.175,  

\bibitem[Krautter et al.(1984)]{1984A&A...137..307K} Krautter, J., et
al.\ 
1984, \aap, 137, 307 

\bibitem[ ]{ }Legget S.K., Hawkins M.R.S., 1989  \mnras\ 238,145 

\bibitem[ ]{ }Margon B., Deutsch E.W.,  1998  \apj\ 498, L61

\bibitem[ ]{ }O'Brien T.J.,  Cohen J., 1998 \apj\ 498, L59

\bibitem[ ]{ }Patterson J., et al. 1998, \pasp\ 110,380

\bibitem[ ]{ }Payne-Gaposchkin C., 1957  The Galactic Novae,  North
Holland 

\bibitem[ ]{ } Rodriguez-Pascual P.M., Gonzalez-Riestra R., Schartel N.,
Wamsteker  W., 1999, \aap\ 139, 183

\bibitem[Saizar \& Ferland(1994)]{1994ApJ...425..755S} Saizar, P., \& 
Ferland, G.~J.\ 1994, \apj, 425, 755 
 
\bibitem[ ]{ }Savage B.D.,  and Mathis J.S., 1979 Ann. Rev. Astron.
Astrophys. 17, 73

\bibitem[Schaefer(2005)]{2005ApJ...621L..53S} Schaefer, B.~E.\ 2005,
\apjl, 
621, L53 

\bibitem[ ]{ }Schaefer B.E., et al., 1992 \apjs\  81,321

\bibitem[ ]{ }Schartel N.,  Skillen I., 1998, in "Ultraviolet
Astrophysics
Beyond the Final Archive " ESA-SP 413, p.735  

\bibitem[ ]{ }Seitter W.C., 1986,  in "RS Ophiuchi",  M.F. Bode (ed.), 
VNU Press,  p.63 

\bibitem[ ]{ }Selvelli P.  Friedjung M.,  2003, \aap\ 401,297

\bibitem[ ]{ }Selvelli P. Gilmozzi R., 2006, in preparation (paper II)

\bibitem[ ]{ }Shabhaz  T., Livio  M., Southwell M., Charles P.A., 1997
\apj\ 484, L59

\bibitem[ ]{ }Shara M.M., Moffat A.F.J., Williams R.E., Cohen J.C.,
1989,
\apj\ 337,720

\bibitem[ ]{ }Shara M.M.,  Zurek D.R., Williams R.E., Prialnik D.,
Gilmozzi R., Moffat A.F.J., 1997 \aj\ 114,258  

\bibitem[Shore et al.(1993)]{1993AJ....106.2408S} Shore, S.~N.,
Sonneborn, 
G., Starrfield, S., Riestra-Gonzalez, R., \& Ake, T.~B.\ 1993, \aj, 106, 
2408 

\bibitem[Smith \& Dhillon(1998)]{1998MNRAS.301..767S} Smith, D.~A., \& 
Dhillon, V.~S.\ 1998, \mnras, 301, 767 

\bibitem[ ]{ }Szkody P., Feinswog L., 1988, \apj\ 334,422

\bibitem[ ]{ }Szkody P., 1994 \aj\ 108,639

\bibitem[ ]{ }Urban Z., 1988, in "Eruptive Phenomena in Stars" Comm.
Konkoly Obs., Acad. Sci. 86, p.359

\bibitem[ ]{ }Vennes S., Thorstensen J.R., Thejll P., Shipman H.L., 
1991,
\apj\ 372, L37

\bibitem[ ]{ }Vogt N., Barrera L.H., Barwig H., Mantel K.H., 1989  in
"Accretion-Powered Compact Binaries",  Mauche C.W. (ed.), Cambridge
Univ. Press, p.391

\bibitem[ ]{ } Wade R.A., Hubeny I., 1998, \apj\ 509, 350

\bibitem[ ]{ }Warner B.,   1995 in "Cataclysmic Variable Stars", 
Cambridge Univ. Press 

\bibitem[ ]{ }Weight A., Evans A., Naylor T., Wood., J., Bode M.F., 1994
\mnras\ 266,761

\bibitem[ ]{ }Webbink R.F., Livio M., Truran J.W., Orio M. (WLTO), 1987,
\apj\   314, 653

\bibitem[Williams et al.(1991)]{1991ApJ...376..721W} Williams, R.~E., 
Hamuy, M., Phillips, M.~M., Heathcote, S.~R., Wells, L., \& Navarrete,
M.\ 
1991, \apj, 376, 721 
 
\bibitem[ ]{ }Young A., Skumanich A., Paylor V., 1988 \apj\ 334, 397  

\bibitem[ ]{ }Young A., Rotler L., Skumanich A., 1991, \apj\ 378, L25 

\end{thebibliography}
\end{document}